\documentclass[aps]{revtex4}
\usepackage{amsfonts}
\newcommand{\C}{\mathbb{C}}

\newcommand{\be}{\nopagebreak[3]\begin{equation}}
\newcommand{\ee}{\end{equation}}
\newcommand{\ba}{\nopagebreak[3]\begin{eqnarray}}
\newcommand{\ea}{\end{eqnarray}}

\begin{document}

\title{Physical effects of the Immirzi parameter}

\author{Alejandro Perez and Carlo Rovelli } 
\email{perez@cpt.univ-mrs.fr,   rovelli@cpt.univ-mrs.fr}
\affiliation{Centre de Physique Theorique de Luminy,
 F-13288 Marseille,\\ European Union},

\vspace{.5cm}

\begin{abstract} \noindent The Immirzi parameter is a constant
appearing in the general relativity action used
as a starting point for the loop quantization of gravity. The
parameter is commonly believed not to show up in the equations of
motion, because it appears in front of a term in the action that
vanishes on shell.  We show that in the presence of fermions,
instead, the Immirzi term in the action does not vanish on shell,
and the Immirzi parameter does appear in the equations of motion.  It
determines the coupling constant of a four-fermion interaction. 
Therefore the Immirzi parameter leads to effects that are observable 
in principle, even independently from nonperturbative quantum gravity.
\end{abstract}

\maketitle
\vskip.5cm

\noindent Connection variables play an important role in the nonperturbative
quantization of general relativity.  Ashtekar realized that the
gravitational field can be effectively described in terms of a
selfdual $SL(2,\C)$ Yang-Mills-like connection and its conjugate
electric field, satisfying appropriate reality conditions \cite{a},
and loop quantum gravity \cite{libro} started as a canonical
quantization of general relativity using these self-dual variables  \cite{lqg}. 
To circumvent difficulties related to the implementation of
the reality conditions in the quantum theory and the noncompatness
of $SL(2,\C)$, the attention has later shifted to a real $SU(2)$
version of the Ashtekar connection, known as the Barbero connection 
\cite{Barbero}.  An action functional that leads directly to both the 
Ashtekar or Barbero formalism is the following, called the Holst action
\cite{Holst}
\be S[e,A]=\frac{1}{16\pi G}\left( \int d^4x\
e e^{a}_{I} e^{b}_{J}\ F^{IJ}_{ab}
-
\frac{1}{\gamma} \int d^4x\
e e^{a}_{I} e^{b}_{J}\ {}^*F^{IJ}_{ab}\right).
\label{action}
\ee Here $I,J\ldots = 0, 1,2,3$ are internal Lorentz indices and $a,b
\ldots = 0, 1,2,3$ are spacetime indices.  The field $e_a^I$ is the
tetrad field, $e$ is its determinant and $e^a_I$ its inverse;
$A_a^{IJ}$ is a Lorentz connection.  $F$ is the curvature of $A$ and
${}^*F$ is its dual, defined by
${}^*F^{IJ}=\frac{1}{2}\epsilon^{IJ}{}_{KL}F^{KL}$.  The coupling
constant $\gamma$ is the Immirzi parameter \cite{Immirzi}.  The choice
$\gamma=i$ leads to the self-dual Ashtekar canonical formalism, while
a real $\gamma$ leads to the $SU(2)$ Barbero connection.

The first term in (\ref{action}) is the familiar tetrad-Palatini
action of general relativity.  The second term does not affect the
equations of motion, for the following reason.  The equation of
motion obtained varying the connection in (\ref{action})  is Cartan's first structure
equation \be
D_{[a}e^I_b{}_{]}=0,
\label{tf}
\ee
where $D_{a}$ is the
covariant derivative defined by $A$.  The solution of this equation is
$A=\omega[e]$, where $\omega[e]$ is the torsion-free spin-connection 
of the tetrad field $e$.  If we thus replace $A$ by $\omega[e]$ in (\ref{action}),
the first term becomes the tetrad expression of the Einstein-Hilbert
action, while the second term is identically zero, due to the Bianchi
identity $R_{[abc]d}=0$.  Stationarity with respect to the variation
of the tetrad yields then the Einstein equations.
Therefore, as it is often stressed, the Immirzi parameter $\gamma$
does not appear to have any effect on the equations of motion.

The parameter $\gamma$, on the other hand, plays an important role in 
loop quantum gravity,
where the spectrum of quantum geometry operators is modulated by
its value.  For instance, the area of a surface and the volume of
a space region are quantized in units of $\gamma \ell_p^2$ and
$\gamma^{3/2} \ell_p^3$ respectively.  Furthermore, the
nonperturbative calculation of the entropy of a black hole appears
to yield a result compatible with Hawking's semi-classical formula
only for a specific value of $\gamma$. See \cite{gamma} for recent
evaluations and references. The role of the Immirzi parameter is
often compared with the role of the $\Theta$ angle in QCD---which
also appears as a constant in front of a term in the action with
no effect on the equations of motion: a parameter that governs
only nonperturbative quantum effects.  See for instance
\cite{RovelliThiemann}. In this letter we point out that this is
in fact not the case in general.

The catch is that the second term in (\ref{action}) vanishes only
when equation (\ref{tf}) is satisfied, but equation (\ref{tf}) is
modified by the presence of fermions (or, more in general, matter
field that couple to the connection, see \cite{Jacobson:1988qt}).
In the presence of a fermion field, (\ref{action}) becomes \be
S[e,A,\psi]=S[e,A] + \frac{i}{2} \int d^4x\ e
\left(\overline\psi\, \gamma^I e_{I}^a D_{a} \psi -
\overline{D_{a}\psi}\, \gamma^I e_{I}^a
 \psi\right), \label{actionf} \ee where $\gamma^I$ are the Dirac
matrices; and (\ref{tf}) becomes \be \label{tff} D_{[a}e^I_{b]}=
{\rm \emph{fermion current}}.  \ee The fermion current acts as a source for
a torsion component in the connection, and the second term in (\ref{action})
doesn't vanish. 

In the following we compute the equations of motion of
(\ref{actionf}), namely the equations of motion of general
relativity coupled with a fermionic field in the presence of a
nontrivial Immirzi parameter.  We solve the equations
for the connection explicitly.  By inserting the solution into the action we
obtain an effective action that contains a four-fermion
interaction. The coupling constant that determines the 
strength of this interaction depends explicitly on the
Immirzi parameter.

\vskip.5cm

Let us start by introducing the tensor 
\be
p_{IJ}{}^{KL}=\frac{1}{2}
(\delta^K_{I}\delta^L_{J}-\delta^L_{I}\delta^K_{J})-\frac{1}{2\gamma}\epsilon_{IJ}{}^{KL}
 \ee and its
inverse 
\be 
p^{-1}{}_{KL}{}^{IJ}=\frac{\gamma^2}{\gamma^2+1}\left(
\frac{1}{2}(\delta^I_{K}\delta^J_{L}-\delta^I_{L}\delta^J_{K})+
\frac{1}{2\gamma}
\epsilon_{KL}{}^{IJ}\right).
\label{p1} 
\ee 
Using this, the action (3) can be written in the
form 
\be 
S[e,A,\psi]=\frac{1}{16\pi G}\int d^4x\ e e^{a}_{I}
e^{b}_{J}\ p^{IJ}{}_{KL} F^{KL}_{ab} + \frac{1}{2}\int d^4x\ e \
\left(i\overline\psi\, \gamma^I e_{I}^a D_{a} \psi\ +\ c.c.
\right), 
\label{actionf2} 
\ee 
and the equation of motion for the
connection reads 
\be p^{IJ}{}_{KL} D_b(e\,e^b_{I} e^a_{J})= 8\pi G
e J_{KL}{}^a, \label{emc} 
\ee 
where the fermion current
$eJ_{KL}{}^a$ is the variation of the fermionic action
with respect to the connection. Recalling that $D_{a}\psi =
\partial_{a}\psi - 1/4\ A_{a}^{KL}\gamma_{K}\gamma_{L}\psi$, and using
the identity $\gamma^A\gamma^{[B}\gamma^{C]}=-i \epsilon^{ABCD}\,
\gamma_5\gamma_D + 2 \eta^{A[B}\gamma^{C]}$, we obtain
\be
J_{KL}{}^a=\frac{1}{8}(2 i e^a{}_{[K}\ j_{\scriptscriptstyle
v}{}_{L]} + e^a_{I}\epsilon^I{}_{KLJ} \ j_{\scriptscriptstyle a}^J
\ +\ c.c.)= \frac{1}{4} e^a_{I}\epsilon^I{}_{KLJ} \ j_{\scriptscriptstyle
a}^J\label{J} 
\ee 
where $j_{\scriptscriptstyle v}^{K} =
\overline{\psi} \gamma^{K}\, \psi$ and $j_{\scriptscriptstyle
a}^{K} = \overline{\psi}\, \gamma_{5}\gamma^{K}\, \psi$ are the 
vector and the axial fermion currents,  and we have used the fact
that they are real.
Using the inverse tensor (\ref{p1}) equation (\ref{emc}) gives
\be
D_b(e\,e^b_{I} e^a_{J})= 8\pi G e \ p^{-1}{}_{IJ}{}^{KL} J_{KL}{}^a.
\label{emc2}
\ee
This equation can be solved for the connection. For this,
we write the connection in the form $A_{a}^{IJ}=\omega[e]_{a}^{IJ}
+ C_{a}{}^{IJ}$, where $\omega[e]$ is the torsion free spin
connection determined by $e$, namely the solution of (2), and $C$
is the torsion. Using this, (\ref{emc2}) gives 
\be 
\label{tf2}
C_{b[K}{}^b e_L{}_{]}^a + C_{[KL]}{}^a =
8\pi G e \ p^{-1}{}_{IJ}{}^{KL} J_{KL}{}^a. 
\ee
Notice that we transform internal and spacetime indices into one
another, using the tetrad field, and preserving the horizontal order
of the indices.  This equation can be solved by contracting the
indices, and then summing terms with cyclical permutation of the
indices: it is easy to verify that the solution is 
\be 
\label{C}
C_a{}^{IJ} = -2\pi G \frac{\gamma}{\gamma^2+1} \left(2 \ e_a^{[I}\
j_{\scriptscriptstyle a}^{J]} - \gamma\ 
e_a^{K}\epsilon_{K}{}^{IJ}{}_L\ 
 j_{\scriptscriptstyle a}^{L} \right). 
 \ee
Notice that the torsion $C$ depends on $\gamma$.

We can now obtain an equivalent action by replacing $A$ with
$\omega[e]+C$ in (3). It is easy to see that the terms linear in the
fermion current are total derivatives, leaving \be S[e,\psi]=S[e]
+ S_{f}[e,\psi] + S_{int}[e,\psi], \ee where the first two terms
are the standard second--order tetrad action of general relativity
with fermions, 
\be 
S[e]+ S_{f}[e,\psi]=\frac{1}{16\pi G} \int
d^4x\ e e^{a}_{I} e^{b}_{J}\ F^{IJ}_{ab}[\omega[e]] \ \ +\ \ i
\int d^4x\ e\ \overline\psi\, \gamma^I e_{I}^a D_{a}[\omega[e]]
\psi 
\ee 
and the interaction term can be obtained by a tedious but
straightforward calculation as 
\be 
S_{int}[e,\psi]= -\frac{3}{2}\pi
G \frac{\gamma^2}{\gamma^2+1} \int d^4x\ e\
(\overline{\psi}\gamma_5\gamma_{A}\psi)\
(\overline{\psi}\gamma_5\gamma^{A}\psi). 
\ee  
This term describes a four-fermion interaction mediated by a 
non-propagating torsion.  An interaction of
this form is well known: it is predicted by the Einstein-Cartan
theory.  The interaction is weak, because it is suppressed by one power of the
Newton constant. It has never been observed, but it is compatible with all present
observations and it might be observed in the future.  Here we see that the coupling 
constant of this interaction
depends on the Immirzi parameter.   In the limit $\gamma\rightarrow \infty$, 
we recover the standard coupling of  the Einstein-Cartan theory.

In summary, general relativity admits the natural formulation
given by the action (3), widely used as a starting point for the
nonperturbative quantization of the theory.  This formulation
includes a four-fermion interaction mediated by the torsion, whose
strength is determined by the Immirzi parameter.  The interaction is present
also on a flat spacetime. The value of the Immirzi parameter is
therefore observable in principle,  independently from its
effect on the nonperturbative quantum theory.  The analogy with
the $\Theta$ angle of QCD is, in this regard, misleading (see also
\cite{alexander}).  The Immirzi parameter is a coupling
constant determining the strength of a four-fermion interaction.

\vskip1cm 

{\em Note added:} In the first version of this work the fermion 
lagrangian was taken to be ${\cal L}_f=ie\overline\psi\, \gamma^I 
e_{I}^a D_{a} \psi$, which yields the effective four-fermion interaction 
\[ S_{int}[e,\psi]= -\frac{3}{2}\pi G \frac{\gamma^2}{\gamma^2+1}
\int d^4x\ e\ \left(j_{\scriptscriptstyle a} \cdot
j_{\scriptscriptstyle a} + j_{\scriptscriptstyle v}\cdot
j_{\scriptscriptstyle v} +\frac{2 i}{\gamma}\ j_{\scriptscriptstyle a} \cdot
j_{\scriptscriptstyle v} \right).\]
Such a fermion lagrangian, however, is not real. This leads to the non-unitarity 
which is manifest in the purely imaginary parity--violating term 
$j_{\scriptscriptstyle a} \cdot j_{\scriptscriptstyle v}$. We
thank I.B. Khriplovich and A.A. Pomeransky, as well
as L. Freidel, for pointing out to us the need of adding the complex
conjugate term to the action. 
A parity--violating four--fermion interaction can nevertheless be still 
obtained without violating unitarity by starting from a
non-minimally-coupled fermion field \cite{laurent}.

\end{document}